\documentclass[12pt, onecolumn]{IEEEtran}

\usepackage[
            left=1in,right=1in,top=1.5 in,bottom=1.5 in,%
            footskip=.25in]{geometry}

\usepackage{algpseudocode}
\algrenewcommand\textproc{}
\algnewcommand{\LineComment}[1]{\State \(\triangleright\) #1}
\usepackage{pifont}
\usepackage[table,xcdraw]{xcolor}
\usepackage{todonotes}
\usepackage{tikz}
\usepackage{float}
\usepackage{newfloat}
\usepackage{color}
\usepackage{verbatim}
\usepackage{amsfonts}
\usepackage{multirow}
\usepackage{bbding}
\usepackage{numprint}
\usepackage{epstopdf}
\usepackage{slashbox}
\usepackage{amsmath}
\usepackage{color}
\usepackage{cite}
\usepackage{mathrsfs}
\usepackage{diagbox}
\usepackage{multirow}
\usepackage{listings}

\usepackage{array}
\newcolumntype{L}[1]{>{\raggedright\let\newline\\\arraybackslash\hspace{0pt}}m{#1}}
\newcolumntype{C}[1]{>{\centering\let\newline\\\arraybackslash\hspace{0pt}}m{#1}}
\newcolumntype{R}[1]{>{\raggedleft\let\newline\\\arraybackslash\hspace{0pt}}m{#1}}

\DeclareMathAlphabet{\mathpzc}{OT1}{pzc}{m}{it}\DeclareFloatingEnvironment[
    fileext=los,
    listname=List of Protocols,
    name=Protocol,
    placement=H,
    within=section,
]{protocol}
\DeclareMathAlphabet{\mathcal}{OMS}{cmsy}{m}{n}

\usepackage{adjustbox}
\usepackage{blindtext}
\usepackage{mathtools}

\usepackage{multirow}
\usepackage{longtable}

\usepackage{etoolbox}
\makeatletter
\patchcmd\@combinedblfloats{\box\@outputbox}{\unvbox\@outputbox}{}{%
  \errmessage{\noexpand\@combinedblfloats could not be patched}%
}%
\makeatother

\usepackage{graphicx}
\usepackage[labelfont=bf, skip=5pt]{caption}
\usepackage{subcaption}

\usepackage[T1]{fontenc}
\usepackage[utf8]{inputenc}
\usepackage{authblk}

\usepackage{url}
\usepackage{algpseudocode}

\usepackage{listings}
\usepackage{tabularx}
\newcolumntype{J}{>{\raggedright\arraybackslash}X}

\usepackage{pgfplots, pgfplotstable}
\pgfplotsset{compat=newest}
\definecolor{carmine}{RGB}{150, 0, 24}
\definecolor{forest}{RGB}{34, 139, 34}
\definecolor{blues4}{RGB}{49, 130, 189}
\pgfplotsset{grid style={dashed,gray}}

\usepackage[normalem]{ulem}
\usepackage[colorlinks=true]{hyperref}

\newif\ifsubmission
\newif\ifcomments
\newif\ifanonymous

\ifdefined\issubmission
\submissiontrue
\fi

\ifdefined\isanonymous
\anonymoustrue
\fi

\ifdefined\iscomments
\commentstrue
\fi

\ifsubmission
\else
\makeatletter
\def\@copyrightspace{\relax}
\makeatother
\fi

\newcommand{\remove}[1]{}

\ifcomments
\newcommand\ghassan[1]{\textcolor{brown}{Ghassan: #1}}
\newcommand\wenting[1]{\textcolor{purple}{Wenting: #1}}
\newcommand\jianliu[1]{\textcolor{blue}{Jian Liu: #1}}
\newcommand\asokan[1]{\textcolor{orange}{Asokan: #1}}
\newcommand\TODO[1]{\textcolor{red}{TODO: #1}}
\else
\newcommand\ghassan[1]{}
\newcommand\wenting[1]{}
\newcommand\jianliu[1]{}
\newcommand\asokan[1]{}
\newcommand\TODO[1]{}
\fi

\newcommand{\Alice}{\text{Alice}}
\newcommand{\Bob}{\text{Bob}}
\newcommand{\TTP}{\text{TTP}}

\begin{document}

\title{Towards Fairness of Cryptocurrency Payments}

\ifanonymous
\else
\author[1]{Jian Liu }
\author[2]{Wenting Li}
\author[2]{Ghassan O. Karame}
\author[1]{N. Asokan}
\affil[1]{Aalto University, Finland}
\affil[2]{NEC Laboratories, Germany}

\fi

\ifsubmission
\IEEEoverridecommandlockouts
\makeatletter\def\@IEEEpubidpullup{9\baselineskip}\makeatother
\IEEEpubid{\parbox{\columnwidth}{Permission to freely reproduce all or part
    of this paper for noncommercial purposes is granted provided that
    copies bear this notice and the full citation on the first
    page. Reproduction for commercial purposes is strictly prohibited
    without the prior written consent of the Internet Society, the
    first-named author (for reproduction of an entire paper only), and
    the author's employer if the paper was prepared within the scope
    of employment.  \\
    NDSS '17, 26 February - 1 March 2017, San Diego, CA, USA\\
    Copyright 2017 Internet Society, ISBN 1-891562-41-X\\
    http://dx.doi.org/10.14722/ndss.2016.23xxx
}
\hspace{\columnsep}\makebox[\columnwidth]{}}
\fi

\maketitle


\begin{abstract}
Motivated by the great success and adoption of Bitcoin, a number of cryptocurrencies such as Litecoin, Dogecoin, and Ethereum are becoming increasingly popular. Although existing blockchain-based cryptocurrency schemes can ensure reasonable security for transactions, they do not consider any notion of fairness.  Fair exchange allows two players to exchange digital ``items'', such as digital signatures, over insecure networks \emph{fairly}, so that either each player gets the other's item, or neither player does. 
Given that blockchain participants typically do not trust each other, enabling fairness in existing cryptocurrencies is an essential but insufficiently explored problem.

In this paper, we explore the solution space for enabling the fair exchange of a cryptocurrency payment for a receipt. We identify the timeliness of an exchange as an important property especially when one of the parties involved in the exchange is resource-constrained. We introduce the notion of \emph{strong timeliness} for a fair exchange protocol and propose two fair payment-for-receipt protocol instantiations that leverage functionality of the blockchain to achieve strong timeliness. We implement both and compare their security and efficiency. 


\subsection*{Keywords}
Cryptocurrency; Bitcoin; Ethereum; Fair Exchange

\remove{
An optimistic fair exchange protocol have an extra feature that guarantees a player can always force a timely and fair termination, without the cooperation of the other player, even in a completely asynchronous network.
However, the original optimistic fair exchange protocol requires a trusted third party in cases where one player attempts to cheat or simply crashes.

A blockchain is a distributed database that maintains a continuously-growing list of data records hardened against tampering and revision. Therefore, it can be used as a permanent and global ledger to track and protect digital signatures which represent items of value, e.g., electronic check, airline ticket or diamond certificate. Sometime, people need to pay to obtain such signatures.

In this paper, we present an optimistic fair exchange protocol to buy or sell blockchain-based signatures without introducing any trusted third party.}


\end{abstract}

\section{Introduction}
\label{sec:intro}

The phenomenal success of Bitcoin~\cite{Nakamoto2009} has fueled innovation in a number of  application domains such as financial payments, smart contracts, and identity management. Recently, a number of blockchain-based solutions have also been proposed in the Internet of Things (IoT) domain. Device-to-device transactions in IoT infrastructures may well make use of cryptocurrency payments in exchange for information or other items of value, without requiring human participation. In scenarios involving resource-constrained devices, it is important to minimize the computation, communication and energy costs of participating in blockchain systems. For example, many blockchain platforms offer the so-called simplified payment verification (SPV) to interface with such devices.


The massive adoption of Bitcoin has truly fueled innovation, and there are currently more than 500 alternate cryptocurencies---most of which are simple variants of Bitcoin. Some of these extensions cannot be deployed without changing the code base of Bitcoin (i.e., via a hard fork). These are referred to as altcoins and require some measures to initiate currency allocation and preserve mining power by leveraging the already established Bitcoin community. 
 
One of the (many) reasons that led to the growing adoption of blockchain-based cryptocurrencies is their premise of low-cost global payments without the need for a bank account, or a cumbersome registration process. However, although existing cryptocurrency schemes can reasonably ensure the security of payments, they do not provide any notion of \emph{fairness}. Given that blockchain participants do not necessarily trust each other, we argue that fairness is an especially important property that should be preserved to ensure the growth of existing cryptocurrency exchanges~\cite{Mayes2014}.

For instance, consider the example where a payer $\Alice$ makes a payment to a payee $\Bob$ in return for an expected good (digital or physical) or a service. This process is unfair towards $\Alice$ if her expectation is not met after $\Bob$ receives the payment. On the other hand, it is unfair towards $\Bob$ if he does provide the service but $\Alice$ later cancels or double-spends the payment. Namely, a fair payment scheme should ensure that $\Bob$ receives the payment if and only if $\Alice$'s expectations are met and vice versa. We can model this as a fair exchange of \textit{payment-for-receipt} where the receipt is a digital signature, which can act as a proxy for a physical/digital good or real-world service.

While there is a wealth of literature on fair exchange in a general setting~\cite{DBLP:journals/jsac/AsokanSW00,Asokan98},
little attention has been paid to the problem of fair exchange involving cryptocurrencies~\cite{Mayes2014}. In this paper, we 
explore the solution space to achieve fair payment-for-receipt for cryptocurrencies. More specifically, we analyze how well known fair exchange techniques can be adapted for use with existing cryptocurrencies, in particular by leveraging functionality from the blockchain. We propose two such protocols and analyze/compare their provisions.

In summary, our contributions in this paper are as follows:
\begin{itemize}
\item We introduce the notion of \emph{strong timeliness} for a fair exchange protocol (cf. Section~\ref{sec:ofe_properties}). We argue that completing the exchanges in a timely and fair manner is an important consideration for fair exchange, especially for resource-constrained devices.
\item We propose \emph{two fair payment-for-receipt protocols for cryptocurrency payments} (cf. Sections~\ref{sec:ttp} and \ref{invasive}) that leverage functionality from the blockchain to meet both fairness and strong timeliness.
  Additionally, we compare these protocols to the proposal by Heilman \emph{et al.}~\cite{HeilmanBG16}.
\item Finally, we implement and evaluate the performance and costs of our constructions~(cf. Section \ref{sec:experimentdesign}).
\end{itemize}

\remove{
The remainder of this paper is organized as follows. In Section~\ref{sec:preliminaries}, we overview the operation of the blockchain and list the requirements that any fair exchange protocol should satisfy. In Section~\ref{timelocking}, we start by describing a possible solution to enable timelock-based fair payments based on the work of Heilman et al.~\cite{HeilmanBG16}. In Section~\ref{sec:ofe}, we present our three instantiations to ensure fair payments based on \emph{(i)} optimistic fair exchange with a stateless blockchain-based TTP, \emph{(ii)} trading non-invasiveness with strong fairness using the blockchain, and finally \emph{(iii)} leveraging witness encryption to achieve fair
exchange in the blockchain. In Section~\ref{sec:experimentdesign}, we evaluate the first two constructs by means of implementation in the Ethereum blockchain. Finally, in Section~\ref{sec:discussion}, we compare the provisions of our proposals and conclude the paper.
}

\section{Blockchain and Smart Contracts}

The notion of blockchain was originally introduced by the well-known hash-based proof-of-work (PoW) mechanism that \emph{confirms} cryptocurrency payments in Bitcoin~\cite{Nakamoto2009}. Bitcoin payments are performed by issuing transactions that transfer Bitcoin coins from the payer to the payee. These entities are called ``peers'', and are referenced in each transaction by means of pseudonyms, denoted by Bitcoin addresses. Each address maps to a unique public/private key pair; these keys are used to transfer the ownership of coins among addresses. Miners are entities that participate in the generation of Bitcoin blocks. These blocks are generated by solving a hash-based PoW scheme; more specifically, miners must find a nonce value that, when hashed with additional
fields (e.g., the Merkle hash of all valid transactions, the hash of the previous block), the result is below a given target value.
If such a nonce is found, miners then include it in a new block thus allowing any entity to verify the PoW. Since each block links to the previously generated block, the Bitcoin \emph{blockchain} grows upon the generation of a new block in the network.

As such, the PoW-based blockchain ensures that all transactions and their order of execution are available to all blockchain nodes, and can be verified by these entities. Consensus by the majority of participating miners is required for every transaction exchanged in the system. This inherently prevents double-spending attacks (where the payer attempts to spend the same coin twice), and ensures the correctness of all transactions confirmed in the blockchain as long as the majority of the network is honest.

To ensure that a payment in a cryptocurrency transaction is definitive, a payee needs to wait until a sufficient number of new blocks have been appended to the block that contains the particular transaction so as to minimize the probability that the block is not part of the eventual consensus. In Bitcoin, this may take up to an hour. In some situations (e.g., low-value transaction), a payee may be willing to accept a transaction as soon as it is broadcast to the network. These are referred to as {\em zero confirmation} transactions, which are fast but carry a risk of payment reversal.

Bitcoin's blockchain fueled innovation, and there are currently more than 500 alternate cryptocurrencies, most of which are simple variants of Bitcoin. Additionally, a number of novel applications have already been devised by exploiting the secure and distributed provisions of the underlying blockchain. Prominent applications include secure timestamping~\cite{Gipp2015}, timed commitment schemes~\cite{Andrychowicz2014}, secure multiparty computations~\cite{Bentov2014}, and smart contracts~\cite{Vitalik2014}.

Smart contracts refer to binding contracts between two or more
parties that are enforced in a decentralized manner by the blockchain nodes without the
need for a centralized enforcer. Smart contracts typically consist of self-contained code that is executed by all blockchain nodes. For example, Ethereum~\cite{Vitalik2014} is a decentralized platform that enables the execution of arbitrary applications (or contracts) on its blockchain. Owing to its support for a Turing-complete language, Ethereum 
offers an easy means for developers to deploy their distributed applications in the form of smart contracts.
%
Ethereum additionally offers its own cryptocurrency \emph{Ether} which is also used as the main fuel to execute the contracts and send transactions. Ether payments are commonly used to cover the costs related to contract execution; these costs are measured by the amount of \emph{gas} they consume. 


\section{Fair Exchange}
\label{sec:ofe_properties}

A two-party exchange usually involves two players who exchange items between themselves. Each player holds an \emph{item} that it wants to contribute to the exchange and an \emph{expectation} about the other player's item it wants to receive in exchange.
Fair exchange is executed between players that do not trust each other; examples include commercial scenarios such as payment-for-receipt, online purchase, digital contract signing, and certified mail. 
A fair exchange protocol must ensure that a malicious player cannot gain any advantage over an honest player.
More specifically, it should satisfy the following requirements~\cite{Asokan98}:  



\begin{itemize}

\item \textbf{Effectiveness:} If both players behave correctly, exchange will eventually happen.

\item \textbf{Fairness:} There are two possible notions of fairness:

\begin{itemize}
\item \textbf{Strong Fairness:} At the time of protocol termination, either both players get what they want, or neither of them does.

\item \textbf{Weak Fairness:} In situations where strong fairness cannot be achieved, an honest player can prove to an (external) arbiter that the other player has received (or can still receive) the item the latter expects.
\end{itemize}

\item \textbf{Timeliness:} Honest players can be certain that the
  protocol will be completed at a certain finite point in time. At
  completion of the protocol, the state of the exchange is final from
  that player's perspective.

\item \textbf{Non-invasiveness:} The protocol should allow the exchange of arbitrary items 
without making any demands on its structure, i.e., the fair exchange protocol does not itself impose any requirement on the form of the items being exchanged. For example, a fair exchange scheme to exchange signatures in any standard digital
  signature algorithm, such RSA or ECDSA, is considered non-invasive. However, if a scheme requires
  anyone who wants to verify the exchanged signatures to access and perform some
  check on the blockchain, then that scheme is
  \emph{invasive}.

\item \textbf{Transaction duration:} The time taken for the exchange
  should be short.
\end{itemize}

\remove{
\begin{description}

\item {\bf Effectiveness: } If $Q$ also behaves correctly, and both $P$ and $Q$ do not want to abandon the exchange, then when the protocol has completed, $P$ has $i_Q$ such that $desc(i_Q)=d_Q$.

\item {\bf Fairness:} two notions of fairness are possible,
\begin{description}
\item {\bf Strong Fairness:} When the protocol has completed, either $P$ has $i_Q$ such that $desc(i_Q) = d_Q$, or $Q$ has gained no additional information about $i_P$.
\item {\bf Weak Fairness:} If strong Fairness cannot be guaranteed, $P$ can prove to an arbiter that $Q$ has received (or can still receive) $i_P$ such that $desc(i_P) = d_P$, without any further intervention from $P$.
\end{description}

\item {\bf Timeliness: } $P$ can be sure that the protocol will be completed at a certain point in time. At completion, the state of the exchange as of that point is either final of any changes to the state will not degrade the level of fairness achieved by $P$ so far.

\item {\bf Non-repudiability:} After an effective exchange (i.e., $P$ has received $i_Q$ at the end of the exchange), $P$ will be able to prove that $i_Q$ originated from $Q$ and $Q$ received $i_p$ .

\end{description}
}

The timeliness requirement defines a fixed point in time at which the protocol will be completed. This property aims to avoid the case where one player in the exchange has to wait indefinitely for the other player to take an action that will determine how the exchange will be concluded (successfully or otherwise). Timeliness is particularly important for resource-constrained IoT devices which cannot afford to be online for long stretches of time or poll indefinitely. One way to achieve the timeliness requirement as stated in~\cite{Asokan98} is to agree on a pre-defined timeout. This is typically a challenging task, since it is difficult to predefine an optimal time point at which the protocol should be completed: a short timeout will result in the exchange failing even when both player are honest (thus harming the effectiveness requirement), whereas a long  timeout is unacceptable for resource-constrained devices with limited battery or bandwidth. 
Ideally, the notion of timeliness should capture the possibility that either player can decide to conclude the exchange \emph{at any point} during the exchange without having to depend on the actions of the other player.
To remedy this, we therefore define a new notion of timeliness, dubbed \emph{strong timeliness}, as follows:

\begin{itemize}
\item \textbf{Strong timeliness:} An honest player can, {\em any}  point in time, choose to complete the protocol. At completion, the state of the exchange is final from that player's perspective. 
\end{itemize}


In this paper, we consider the ``payment-for-receipt'', where an entity, $\Alice$, makes a digital payment to another entity, $\Bob$, in order to get
a receipt for the payment in the form of a digital signature. 
%
%
\emph{Our goal is to explore the solution space for integrating a fair exchange of payment-for-receipt into existing cryptocurrency payment schemes.} (Hereafter referred to as ``fair payments'' for the sake of brevity.)

\section{Fair Payments with Fixed Timeouts}
\label{timelocking}

Recently, a number of fair blockchain-based protocols have been proposed~\cite{Andrychowicz2014, Bentov2014, HeilmanBG16, Kumaresan:2016:ISC:2976749.2978421}. All of them rely on the use of \emph{fixed timeouts} to ensure fairness. In this section, we discuss why the use of fixed timeouts can negatively impact fairness guarantees, and possibly even the overall security of the protocol. As an example, we use a protocol by Heilman et al.~\cite{HeilmanBG16} that uses timeout-based fair payments using an intermediary to improve the anonymity of cryptocurrency payments.
Here, a user $\Alice$ wants to fairly exchange a cryptocurrency payment for a
voucher from an intermediary $\Bob$. The voucher
is in fact a blind signature; $\Alice$ will unblind the voucher and send
it privately to an anonymous payee who can exchange it with $\Bob$ in
such a way that $\Bob$ cannot link $\Alice$ and the payee~\cite{HeilmanBG16}.

This is achieved without relying on any external entity through the use of blockchain-based script and using 
a fixed, predefined, timeout to implement a timely fair exchange. 
First, $\Alice$ generates a transaction that enables
her to pay a pre-defined amount to $\Bob$ under the condition that $\Bob$
must publish a valid signature on a message within a certain time window;
The output of this transaction will become an input in one of the following two blockchain transactions:
\begin{enumerate}
\item A transaction which is signed by $\Bob$ and contains a valid signature on the requested message (i.e., exchange is successful and fair).
\item A transaction which is signed by $\Alice$ and the time window has expired. (i.e., exchange fails and the money reverts to
  $\Alice$).
\end{enumerate}

The condition is fulfilled if $\Bob$ posts a transaction that contains a valid signature 
and the promised payment is transferred from $\Alice$ to $\Bob$.
If $\Bob$ does not publish a signature within the time window, $\Alice$ can sign and post a transaction that returns the promised payment amount back to herself. 
All transactions are broadcast to the blockchain network, thus allowing all blockchain miners to verify  whether the payment conditions have been met, and reach consensus on the state of the exchange.


This protocol ensures a fair exchange between $\Alice$ and $\Bob$,
i.e., prevents $\Alice$ from double-spending her payment, and enables $\Bob$ to spend $\Alice$'s payment only if $\Bob$ has published his signature.
We now analyze this protocol in relation to the requirements from Section~\ref{sec:ofe_properties}:

\begin{itemize}

\item {\bf Effectiveness:} If the timeout is too short, there may be not enough time for Bob's transaction to be confirmed in the blockchain. Namely, the miners will refuse to confirm that transaction after the timeout has passed. In this case, the effectiveness of the fair exchange cannot be guaranteed since the exchange fails because of the timeout even when both parties behave correctly.

\item {\bf Fairness:} The protocol does not ensure strong fairness since it is possible that the timeout is reached after $\Bob$ broadcasts his transaction, but before it is confirmed in the blockchain. For example, the adversary may mount a denial-of-service attack against $\Bob$ to throttle its network connectivity~\cite{Gervais:2015:TDB:2810103.2813655}. In this case, $\Alice$ might receive the signature without $\Bob$ receiving the payment. 
However, the protocol satisfies weak fairness, since $\Bob$ can prove to an (external) arbiter that his signature on Alice's requested message has indeed been revealed to the public. Note that this renders the anonymous payment scheme of~\cite{HeilmanBG16} insecure!

\item {\bf Timeliness:} This protocol satisfies weak timeliness but not strong timeliness, since once the Bob's transaction is confirmed, $\Alice$ cannot decide to complete the exchange any sooner than the specific timeout.

\item {\bf Non-invasiveness:} The protocol is non-invasive because it does not impose any specific structure on Bob's signature.

\item {\bf Transaction duration:} $\Bob$ needs to wait for a period such that enough blocks has been appended to the blockchain after its own transaction appears there. $\Alice$ has to wait at most till the timeout. 
\end{itemize}

\section{Optimistic Fair Payments} 
\label{sec:ofe}


\subsection{Optimistic Fair Exchange}

Optimistic fair exchange (OFE) protocols were first proposed by Asokan
et al.~~\cite{DBLP:journals/jsac/AsokanSW00,Asokan98};
their protocol relies on the presence of a trusted third party ($\TTP$) but only in an \emph{optimistic} fashion: $\TTP$ is only required when one player attempts to cheat or simply crashes. In the common case where $\Alice$ and $\Bob$ are honest and behave correctly, $\TTP$ need not to be involved.

Optimistic fair exchange consists of an exchange protocol (protocol $\mathsf{exchange}$) and two recovery protocols (protocol $\mathsf{abort}$ and protocol $\mathsf{resolve}$). 
First, both players agree on what needs to be exchanged and which third party to use in case of an exception.
Such an ``agreement'' is informal: it has no validity outside the context of the protocol.
Then, one player (e.g., $\Alice$) sends a verifiable encryption ($c_A$) of her item ($i_A$) and her expectation about $\Bob$'s item ($e_A$).
The verifiable encryption enables any entity to verify the validity of  $i_A$ (without the need for decrypting the message); and can be decrypted only by the trusted third party $\TTP$.
$\Bob$ first verifies $i_A$, constructs  an encryption $c_B$ of $(i_B, e_B)$, and decides similarly whether to send it to $\Alice$.

If $\Bob$ does not send $c_B$, $\Alice$ can abort the protocol at any point in time by initiating protocol $\mathsf{abort}$ with $\TTP$ which issues an abort token. In this case, the exchange is unsuccessful but fair: neither player receives any additional information about each other's item.
If $\Bob$ sends $c_B$, and $\Alice$ has not decided to abort, she verifies the validity of $i_B$ and decides whether to send $i_A$ to $\Bob$. While waiting for $i_A$, $\Bob$ can initiate protocol $\mathsf{resolve}$ at any time by sending $(c_A, i_B)$ to $\TTP$. $\TTP$ will decrypt $c_A$ to get $(i_A, e_A)$ and return $i_A$ to $\Bob$ if $i_B$ meets $e_A$ and it has not previously issued an $abort$ token for this particular exchange. 
If a transaction was previously aborted, $\TTP$ will not agree to resolve it. Similarly it will not agree to abort a transaction that had already been resolved.
$\Alice$ can run $\mathsf{resolve}$ in the same way while waiting for $i_B$ from $\Bob$.
This is a general fair exchange protocol that can support ``items'' in the form signatures in standard signature schemes. It requires $\TTP$ to keep state for every aborted or resolved transaction.

\subsection{Blockchain-based OFE with a Stateless $\TTP$}
\label{sec:ttp}

We now extend the above OFE protocol by making use of a blockchain to avoid the need for $\TTP$ to maintain state.
$\Alice$ can abort the exchange by publishing an $\mathsf{abort}$ transaction to the blockchain instead of sending an $\mathsf{abort}$ message to $\TTP$.
Thus $\TTP$ only needs to support the $\mathsf{resolve}$ protocol.
It does not need to keep any state w.r.t. the protocol execution since all needed state information is recorded in the blockchain.
We implemented this variant of OFE using Ethereum's smart contracts as shown in Figure~\ref{BlockchainOFE}.
Note that when $\TTP$ recovers item $i_A$ in response to a $\mathsf{resolve}$ request from $\Bob$, it needs to save $\Bob$'s item $i_B$ so that any subsequent $\mathsf{abort}$ from $\Alice$ can be answered correctly by the smart contract without violating $\Alice$'s fairness.
Therefore, $\TTP$ will ask the smart contract to save $i_B$ during $\Bob$'s invocation of $\mathsf{resolve}$.
\begin{figure}[t]
\begin{adjustbox}{minipage=0.99\linewidth,fbox,center}
\begin{algorithmic}
    \Function{abort}{exchange id $id_{ex}$}
	\If{entry of $id_{ex}$ exists}
      \If{sender is the originator and the retrieved entry is a resolved item $i_B$}
          \State return $i_B$ to the sender
	  \EndIf
	\Else
      \State add an entry of $id_{ex}$ with an $\mathsf{abort}$ token
	\EndIf
    \EndFunction
\\
    \Function{resolve}{exchange id $id_{ex}$, optional resolved item $i_B$}
      \If{sender is $\TTP$}
        \If{entry of $id_{ex}$ exists and the retrieved entry is an $\mathsf{abort}$ token}
          \State return \textit{aborted}
        \Else
		  \State add an entry of $id_{ex}$ with the optional resolved item $i_B$
          \State return \textit{$\neg$~aborted}
        \EndIf
	  \EndIf
    \EndFunction			
\end{algorithmic}
\end{adjustbox}
    \caption{Smart contract for blockchain-based OFE to assist $\mathsf{abort}$
             and $\mathsf{resolve}$ procedures in order to keep $\TTP$ stateless.}
\label{BlockchainOFE}
\end{figure}

We can easily build a fair payment protocol based on this blockchain-based OFE, by having $i_A$ be a signature corresponding to a payment message in a cryptocurrency scheme.
First, we consider zero confirmation payments: the two players exchange payment for a receipt but do not wait for the payment to be confirmed in the blockchain.
We now analyze this protocol w.r.t. the properties listed in Section~\ref{sec:ofe_properties}:

\begin{itemize}
\item {\bf Effectiveness:} Effectiveness is guaranteed if both players behave correctly, since Alice will get Bob's signature immediately after the OFE and her payment will be eventually confirmed.
\item {\bf Fairness:} Once the OFE completes, $\Alice$ receives $i_B$ and $\Bob$ receives $i_A$. 
However, $\Alice$ can double-spend the money associated with $i_A$ after the completion of OFE, thus invalidating strong fairness. $\Bob$ can however prove this misbehavior to an arbiter by showing $i_A$. So, this scheme only satisfies the weak fairness property.
\item {\bf Timeliness:} Strong timeliness is inherited from classical OFE: either player can invoke protocol $\mathsf{resolve}$ at any point if they have received the other player's verifiable encryption ($c_A$ or $c_B$). $\Alice$ can attempt to $\mathsf{abort}$ at any time. In all cases, the protocol is guaranteed to terminate fairly.
\item {\bf Non-invasiveness:} The signature $i_B$ is non-invasive since it can be any signature in any form. 
\item {\bf Transaction duration:} Since the transactions are zero confirmation, they can complete fast during optimistic execution (no need to wait any blocks confirmed on the blockchain).
\end{itemize}

This variant can be upgraded from zero confirmation to full confirmation by borrowing the approach of Mayes et al\cite{Mayes2014} to  require that $\Bob$ and $\TTP$ check if $i_A$ is confirmed on the blockchain as follows. After getting $i_A$ from $\Alice$, $\Bob$ broadcasts it and waits for it to be confirmed on the blockchain before sending $i_B$ to $\Alice$. When $\Bob$ asks $\TTP$ to resolve, $\TTP$ similarly broadcasts $i_A$ and waits for it to be confirmed on the blockchain before storing $i_B$. When it resolves for $\Alice$, it returns $i_B$ only after $i_A$ is confirmed on the blockchain. If $i_A$ was double spent before being confirmed, $\TTP$ will treat it as though $\Alice$ aborted the protocol. This full confirmation variant achieves strong fairness but at the expense of longer transaction duration.

\section{Fair Payments of Blockchain-based Signatures}
\label{invasive}



We now describe a variant that dispenses with the need for $\TTP$
altogether but at the expense of making the signature invasive.
Our proposal unfolds as follows. $\Alice$ first constructs the message to be signed, and uses it to create a transaction with an output of some amount of digital money that is spendable in one of the following two transactions:
\begin{enumerate}
\item A transaction that is signed by $\Bob$ and contains a valid signature on the requested message;
\item An $\mathsf{abort}$ transaction that is signed by $\Alice$.
\end{enumerate}

Notice that there are no time window/constraints in Alice's transaction, and it can trigger either of the above two transactions, depending on which one is confirmed in the blockchain first.
Recall that if both are broadcast, only one of them will eventually be confirmed (since they conflict with one another).
This protocol is invasive since  $\Bob$'s signature is only valid if it is stored on the blockchain, a.k.a. blockchain-based signature. 
Namely, a verifier must not only check that the signature is (cryptographically) valid, but also that it is confirmed in the blockchain.

This protocol can be fully deployed as an Ethereum smart contract without the need for $\TTP$. In this case, $\Alice$ will first send a deposit to the contract; the contract will forward the deposit either to $\Bob$ or back to $\Alice$, depending on whether it receives Bob's signature or Alice's $\mathsf{abort}$ first (respectively).
An example of such contract functions is sketched in Figure~\ref{BlockchainSig}.

\begin{figure}[t]
\begin{adjustbox}{minipage=0.99\linewidth,fbox,center}
\begin{algorithmic}
    \Function{initExchange}{payment $T_{pay}[v]$, expected item $m$ and recipient}
        \If{state is UNINITIALIZED and contract has received the payment with value $v$}
            \State record originator, recipient and $m$
            \State switch state to INITIALIZED
        \EndIf
    \EndFunction
	\Function{abort}{}
        \If{state is INITIALIZED and the message is sent by the originator}
			\State refund $v$ to the originator
			\State clear up storage and switch state to UNINITIALIZED
        \EndIf
	\EndFunction
    \Function{resolve}{signature on $m$}
         \If{state is INITIALIZED and the message is sent by the recipient}
            \If{signature on $m$ is valid}
				\State send $v$ to recipient and the signature to originator
                \State clear up storage and switch state to UNINITIALIZED
            \EndIf
        \EndIf
    \EndFunction			
\end{algorithmic}
\end{adjustbox}
\caption{Smart contract for fair payment of blockchain-based signature.}
\label{BlockchainSig}
\end{figure}

Our extension ensures the following properties:
\begin{itemize}
\item {\bf Effectiveness:} If both players behave correctly, $\Bob$ will receive the payment by publishing a signature, and $\Alice$ will obtain her desired receipt when the signature has been confirmed on the blockchain.
\item {\bf Fairness:} $\Bob$ can only receive the payment when his signature is confirmed on the blockchain. Similarly, once it is confirmed, $\Alice$ has a valid receipt. Ether the signature or the $\mathsf{abort}$  will be confirmed, but not both.
\item {\bf Timeliness:} The protocol terminates after either the signature or the $\mathsf{abort}$ is confirmed. $\Alice$ can choose to wait for the signature to be confirmed or issue an $\mathsf{abort}$. In the former case, the exchange is successfully completed. In the latter case, $\Alice$ cannot gain any advantage since the signature is valid only if it is confirmed in the blockchain. Similarly $\Bob$ can either issue a signature and wait for it to be confirmed, or simply walk away. In either case, the state of the exchange is final.
\item {\bf Non-invasiveness:} Clearly, the signature is invasive since it is only valid when it is confirmed in the blockchain.
\item {\bf Transaction duration:} The transaction duration is long because both parties need to wait for either Bob's signature or Alice's $\mathsf{abort}$ to be confirmed on the blockchain.
\end{itemize}


\section{Experimental Evaluation}
\label{sec:experimentdesign}

We now describe and evaluate our Ethereum-based implementation of the contracts depicted in Figure~\ref{BlockchainOFE} (fair payment with blockchain-based OFE) and Figure~\ref{BlockchainSig} (fair payment with blockchain-based signature).




\subsection{Implementation Setup}

We assigned an Ethereum node to each entity (e.g., $\Alice$, $\Bob$, $\TTP$).
These nodes are connected to a private Ethereum network (that is equipped with private mining functionality) with a bandwidth limit of 100Mbps. We deployed the mining node and $\TTP$ on two servers both with 24-core Intel Xeon E5-2640 and 32GB of RAM. In our testbed, $\Alice$ and $\Bob$ reside on two machines equipped with 4-core Intel i5-6500 with 8GB of RAM and 8-core Intel Xeon E3-1230 with 16GB of RAM, respectively. In our implementation, these entities prepare and send the transactions to the blockchain using the Javascript library web3.js. This library interfaces the Ethereum nodes through its RPC calls.
In the blockchain-based OFE instantiation, we implement OFE computation and communication using GoLang and C.
%
%
We use the ECDSA signature scheme, which is directly supported by Ethereum contracts.
We use the verifiable encryption scheme in \cite{Asokan98} implemented with cryptographic library GMP \cite{libgmp} in C. We preset and fix the difficulty of our private Ethereum testnet in the code and the genesis block so that the block generation time is around 5 seconds.

In our experiments, we measured the gas and time consumption for the following procedures: $\mathsf{deploy}$, $\mathsf{optimistic~completion}$, $\mathsf{abort}$, and $\TTP$ $\mathsf{resolve}$.
$\mathsf{Deploy}$ refers to deploying the smart contract into the blockchain.
In blockchain-based OFE, the smart contract is deployed by $\TTP$ to manage its state.
$\mathsf{Optimistic~completion}$ refers to the successful completion of the exchange without invoking  $\mathsf{resolve}$ or  $\mathsf{abort}$.
Finally, the contract is triggered by $\Alice$ for abort and by $\TTP$ for resolve.
We only consider the $\mathsf{resolve}$ protocol under the assumption that the exchange has not been aborted.

To measure gas consumption, we observe the difference in the account balance before and after invoking the contract, and we convert this amount to the amount of gas according to our fixed gas price.
We measure the {\em eclipsed time} starting from the initial contract invocation until the entities receive the notifications from the Ethereum network. In our evaluation, each
time measurement is averaged over 10 independent executions of the fair exchange protocol; where appropriate, we also report the corresponding 95\% confidence intervals.

\begin{table}[tb]
\centering
\begin{tabular}{|c|c|c|c|}
\hline
\multicolumn{2}{|c|}{\multirow{2}{*}{\backslashbox{\small Actions}{\small Protocols}}} &\small with blockchain-based OFE &\small with blockchain-based signature     \\
\multicolumn{2}{|c|}{}                                                                 &\small (Section~\ref{sec:ttp})            &\small  (Section~\ref{invasive})      \\ \hline
    \multicolumn{2}{|c|}{$\mathsf{deploy}$}           & 537,783 		     & 645,900  \\ \hline
    \multirow{2}{*}{ \parbox{2.5cm}{$\mathsf{optimistic~completion}$} }   & $\Alice$ 			& \multirow{2}{*}{0} & 126,457 \\ \cline{2-2} \cline{4-4}
    											   & $\Bob$ 			& 					  & 27,935   \\ \hline
    \multicolumn{2}{|c|}{$\mathsf{abort}$}           & 67,574  			&  33,746  \\ \hline
    \multicolumn{2}{|c|}{$\TTP$ $\mathsf{resolve}$}     & 132,600 			& -- \\ \hline
\end{tabular}
\caption{Gas consumption in Ethereum contracts to perform each action of blockchain-based fair payment protocols.}
\label{tb:gasConsumption}
\end{table}

\subsection{Evaluation Results}

\subsubsection{Gas consumption}

Our evaluation results are shown in Table~\ref{tb:gasConsumption}.
We first observe that for both contracts, contract deployment consumes the most amount of gas since the process of creating contracts and storing data in the blockchain are expensive operations in Ethereum~\cite{EthereumYellowPaper}.
As described earlier, the blockchain-based OFE variant does not need to involve the blockchain at all during optimistic exchanges.
Therefore the gas consumption for an optimistically concluded fair payment is zero.
We contrast this with the blockchain-based Signature variant (Section~\ref{invasive}),
 which requires 126,457 gas from $\Alice$ in order to initiate the exchange protocol and 27,935 gas from $\Bob$ to complete it.
 The large overhead incurred on $\Alice$ here is mainly caused by storing exchange contract parameters in the blockchain during contract initialization.
Notice that $\mathsf{abort}$ requires considerably more gas in blockchain-based OFE when compared to the blockchain-based Signature (Section~\ref{invasive}).
This is due to the fact that the contract may spend more gas in order to transmit the previous resolved item (if any).
Similarly, $\TTP$ $\mathsf{resolve}$ 
potentially needs to store resolved items in the contract (cf. Section~\ref{sec:ttp}) -- which incurs additional gas consumption.

\subsubsection{Time consumption}


Table~\ref{tb:timeConsumption} shows the measured eclipsed time.
We observe that the contract invocation process
is rather time-consuming; for instance, the protocol initialization procedure by $\Alice$ in blockchain-based Signature (Section~\ref{invasive}) consumes around 4 seconds.
We contrast this with 277 milliseconds which is required for the completion of the blockchain-based OFE protocol.
The latter is almost 14 times faster in spite of the reliance on verifiable encryption, due to the fact that the blockchain needs to generate a block in order to include the transactions. Recall that the average block generation
 time in our private Ethereum network is tuned to be around 5 seconds.

The time execution of the remaining operations is comparable in both protocols, which is around 4 seconds. This value largely depends on block generation time of the blockchain network.
Notice that the width of the confidence interval corresponds to the variation of block generation times exhibited in Ethereum.

\subsubsection{Summary}

Given our findings, we conclude that the fair payment protocol based on blockchain-based OFE
is more cost- and time-effective than its counterpart based on blockchain-based Signatures (Section~\ref{invasive}) when the protocol is executed without exceptions.

In the case where an exception occurs, both protocols incur comparable costs and time overhead. Namely, since any transaction can only take effect once they are confirmed in the blockchain (i.e., every 12
seconds in the Ethereum public blockchain), the reliance on the blockchain in $\mathsf{abort}$ and $\mathsf{resolve}$ protocol incurs considerable time delays.


\begin{table}[tb]
\centering
\begin{tabular}{|c|c|c|c|}
\hline
\multicolumn{2}{|c|}{\multirow{2}{*}{\backslashbox{\small Actions}{\small Protocols}}} &\small with blockchain-based OFE  &\small with blockchain-based signature     \\
\multicolumn{2}{|c|}{}                                                                 &\small (Section~\ref{sec:ttp})            &\small  (Section~\ref{invasive})      \\ \hline
	\multirow{2}{*}{\parbox{2.5cm}{$\mathsf{optimistic~completion}$}} & $\Alice$ & \multirow{2}{*}{$277.0 \pm 9.2$}  & $3,831.0 \pm 973.2$ \\ \cline{2-2} \cline{4-4}
							   & $\Bob$ &                                   & $4,195.8 \pm 1,077.3$ \\ \hline
    \multicolumn{2}{|c|}{$\mathsf{abort}$}       		& $3,432.4 \pm 1,000.3$  	& $4,301.5 \pm 1,305.6$ \\ \hline
    \multicolumn{2}{|c|}{$\TTP$ $\mathsf{resolve}$}      & $3,773 \pm 902.5$ 		& -- \\ \hline
\end{tabular}
\caption{Eclipsed time in milliseconds with 95\% confidence interval to perform each action of blockchain-based fair payment protocols.}
\label{tb:timeConsumption}
\end{table}

\section{Comparison and Outlook}
\label{sec:discussion}

\begin{table}[]
\centering
\begin{tabular}{|c|c|c|c|c|}
\hline
\multirow{2}{*}{\backslashbox{\small Requirements}{\small Types}}    &
                                                                       \multirow{2}{*}{\begin{tabular}[c]{@{}c@{}}with fixed timeout\\ (Section~\ref{timelocking})\end{tabular}} & \multicolumn{2}{c|}{\begin{tabular}[c]{@{}c@{}}with blockchain-based OFE\\ (Section~\ref{sec:ttp})\end{tabular}} & \multirow{2}{*}{\begin{tabular}[c]{@{}c@{}}with blockchain-based signature\\ (Section~\ref{invasive})\end{tabular}} \\ \cline{3-4}
                        &                                                                                                                       & zero-confirm                                      & full-confirm                                     &                                                                                                 \\ \hline
Fairness                &         weak                                                                                                &     weak                                              &    strong                                         & strong                                                                                                \\ \hline
Timeliness              &        weak                                                                                                &     strong                                              &       strong                                           &       sting                                                                                          \\ \hline
Effectiveness           &           ?                                                                                                  &    $\surd$                                               &      $\surd$                                            &      $\surd$                                                                                           \\ \hline
Non-invasiveness        &      $\surd$                                                                                         &     $\surd$                                             &   $\surd$                                                  &        X                                                                                         \\ \hline
No $\TTP$               &     $\surd$                                                                                             &       X                                            &           X                                       &                $\surd$                                                                                 \\ \hline
Duration of transaction &      long                                                                                             &  short                                                   &     long                                             &      long                                                                                           \\ \hline
\end{tabular}
\caption{Comparisons of different blockchain-based fair payment protocols.}
\label{comparisons}
\end{table}


In this paper, we explored the solution space to realize fair exchange within cryptocurrencies. To this end, we proposed {two fair payment-for-receipt protocols for cryptocurrency payments} (cf. Sections~\ref{sec:ttp} and \ref{invasive}) that leverage functionality from the blockchain to meet both fairness and strong timeliness.
A systematic comparison between our proposals is shown in Table~\ref{comparisons}.

Our findings suggest that the first scheme cannot satisfy the strong timeliness property, and as such can only guarantee weak fairness. Furthermore, choosing a short-timeout here can harm the effectiveness of this construct.
In this respect, the constructs based on blockchain-based OFE  (Section~\ref{sec:ttp}) and blockchain-based  signature (Section~\ref{invasive}) establish the strongest tradeoffs between performance and provisions. 
Namely, our performance evaluation shows that the blockchain-based OFE option is more efficient when the exchange concludes optimistically. 
As such, it seems to be ideal in those scenarios where only weak fairness is sufficient or when non-invasiveness is required. 
Otherwise, we recommend the reliance on the blockchain-based signature option.

\remove{
None of the variants we examined can simultaneously possess all desirable properties (strong fairness, strong timeliness, effectiveness, non-invasiveness, no use of $\TTP$ and short transaction duration). 
Designing a fair payment mechanism that does possess all of these properties remains an open question. 
A potential option is to leverage the recent development of {\em witness encryption}~\cite{Garg2013}, which is associated with a specific NP-relation $(x, w) \in R$, where $x$ is a ``statement'' and $w$ is a ``witness''.
It allows the sender to encrypt a message $m$ with respect to statement $x$ as $c \leftarrow Enc_R(x,m)$. 
Any receiver who knows the witness $w$ that satisfies $(x,w) \in R$ can decrypt this ciphertext $c$ as $m = Dec_R(c,w)$. 
So, we can improve the Invasive Signature (Section~\ref{invasive}) approach by having $\Bob$ broadcast a non-invasive signature encrypted under the witness encryption, and the witness for decryption is that the blockchain contains $T_{\sigma}$ but does not contain $T_{\mathsf{abort}}$. 
We did not implement this construction since all known constructions of witness encryption are penalizing in terms of performance.
}

Privacy has not been considered as a requirement for fair exchange protocols. 
However, there are a number of scenarios where privacy considerations play a paramount role. 
For example, the message to be signed may contain some important information about $\Alice$ that cannot be revealed. 
In all of the three constructions, the contents of the signature can be seen by the public; there are, however, a number of techniques that can be used to protect the contents of signatures. 
For instance, one can improve the blockchain-based OFE (Section~\ref{sec:ttp}) protocol by having $\TTP$ send a verifiable encryption of $i_B$ to the resolve contract---thus preserving the privacy of $\Alice$.



In conclusion, while there are no ``bullet-proof'' solutions that simultaneously achieve all desirable properties discussed above
we observe that a number of trade-offs exist within the solution space to sacrifice one property in order to achieve the rest. 
Depending on the application scope, this might already offer a differentiator and stronger value proposition for existing cryptocurrencies. 
We therefore hope that our findings motivate further research in this largely unexplored area.

%



\raggedright
\bibliographystyle{unsrt}
\bibliography{sigproc}

\begin{thebibliography}{10}

\bibitem{Nakamoto2009}
Satoshi Nakamoto.
\newblock Bitcoin: A peer-to-peer electronic cash system, 2009.
\newblock \url{http://www. bitcoin.org/bitcoin.pdf}.

\bibitem{Mayes2014}
D.~Jayasinghe, K.~Markantonakis, and K.~Mayes.
\newblock Optimistic fair-exchange with anonymity for bitcoin users.
\newblock In {\em e-Business Engineering (ICEBE), 2014 IEEE 11th International
  Conference on}, pages 44--51, Nov 2014.
\newblock
  \url{http://ieeexplore.ieee.org/document/6982058/?reload=true&arnumber=6982058}.

\bibitem{DBLP:journals/jsac/AsokanSW00}
N.~Asokan, Victor Shoup, and Michael Waidner.
\newblock Optimistic fair exchange of digital signatures.
\newblock {\em {IEEE} Journal on Selected Areas in Communications},
  18(4):593--610, 2000.
\newblock \url{http://dx.doi.org/10.1109/49.839935}.

\bibitem{Asokan98}
N.~Asokan.
\newblock {\em Fairness in Electronic Commerce}.
\newblock PhD thesis, University of Waterloo, 1998.
\newblock
  \url{https://uwspace.uwaterloo.ca/bitstream/handle/10012/292/NQ32811.pdf}.

\bibitem{HeilmanBG16}
Ethan Heilman, Foteini Baldimtsi, and Sharon Goldberg.
\newblock Blindly signed contracts: Anonymous on-blockchain and off-blockchain
  bitcoin transactions.
\newblock In Jeremy Clark et~al., editors, {\em Financial Cryptography and Data
  Security - {FC} 2016 International Workshops, BITCOIN, VOTING, and WAHC,
  Christ Church, Barbados, February 26, 2016, Revised Selected Papers}, volume
  9604 of {\em Lecture Notes in Computer Science}, pages 43--60. Springer,
  2016.
\newblock \url{http://eprint.iacr.org/2016/056}.

\bibitem{Gipp2015}
Bela Gipp, Norman Meuschke, and Andr{\'{e}} Gernandt.
\newblock Decentralized trusted timestamping using the crypto currency bitcoin.
\newblock {\em CoRR}, abs/1502.04015, 2015.
\newblock \url{http://arxiv.org/abs/1502.04015}.

\bibitem{Andrychowicz2014}
Marcin Andrychowicz, Stefan Dziembowski, Daniel Malinowski, and Lukasz Mazurek.
\newblock Secure multiparty computations on bitcoin.
\newblock In {\em Proceedings of the 2014 IEEE Symposium on Security and
  Privacy}, SP '14, pages 443--458, Washington, DC, USA, 2014. IEEE Computer
  Society.
\newblock \url{http://dx.doi.org/10.1109/SP.2014.35}.

\bibitem{Bentov2014}
Iddo Bentov and Ranjit Kumaresan.
\newblock How to use bitcoin to design fair protocols.
\newblock In Juan~A. Garay and Rosario Gennaro, editors, {\em Advances in
  Cryptology -- CRYPTO 2014: 34th Annual Cryptology Conference, Santa Barbara,
  CA, USA, August 17-21, 2014, Proceedings, Part II}, pages 421--439, Berlin,
  Heidelberg, 2014. Springer Berlin Heidelberg.
\newblock \url{http://dx.doi.org/10.1007/978-3-662-44381-1_24}.

\bibitem{Vitalik2014}
Vitalik. Buterin.
\newblock A next-generation smart contract and decentralized application
  platform, 2014.
\newblock \url{https://github.com/ethereum/wiki/wiki/White-Paper}.

\bibitem{Kumaresan:2016:ISC:2976749.2978421}
Ranjit Kumaresan, Vinod Vaikuntanathan, and Prashant~Nalini Vasudevan.
\newblock Improvements to secure computation with penalties.
\newblock In {\em Proceedings of the 2016 ACM SIGSAC Conference on Computer and
  Communications Security}, CCS '16, pages 406--417, New York, NY, USA, 2016.
  ACM.
\newblock \url{http://doi.acm.org/10.1145/2976749.2978421}.

\bibitem{Gervais:2015:TDB:2810103.2813655}
Arthur Gervais, Hubert Ritzdorf, Ghassan~O. Karame, and Srdjan Capkun.
\newblock Tampering with the delivery of blocks and transactions in bitcoin.
\newblock In {\em Proceedings of the 22Nd ACM SIGSAC Conference on Computer and
  Communications Security}, CCS '15, pages 692--705, New York, NY, USA, 2015.
  ACM.
\newblock \url{http://doi.acm.org/10.1145/2810103.2813655}.

\bibitem{libgmp}
GMP.
\newblock The gnu multiple precision arithmetic library.
\newblock \url{https://gmplib.org/}.
\newblock Accessed: 2016-07-26.

\bibitem{EthereumYellowPaper}
Wood Gavin.
\newblock Ethereum: A secure decentralised generalised transaction ledger.,
  2014.
\newblock
  \url{http://bitcoinaffiliatelist.com/wp-content/uploads/ethereum.pdf}.

\end{thebibliography}

\ifsubmission
\else
\fi

\end{document}